\documentclass{sig-alternative-10pt}
 \pdfoutput=1
\usepackage[T1]{fontenc}
\usepackage{lmodern}
\usepackage[utf8]{inputenc}
\usepackage[colorlinks=False]{hyperref}
\usepackage{multirow}

\usepackage{enumitem}

\usepackage{float}
\begin{document}
%

\title{WhoTracks.Me: Shedding light on the opaque world of online tracking}
%
%
%
%
%

\numberofauthors{4} 
%
\author{
%
%
\alignauthor
Arjaldo Karaj\\
       \email{arjaldo@cliqz.com}
\alignauthor
Sam Macbeth\\
       \email{sam@cliqz.com}
\alignauthor
Rémi Berson\\
       \email{remi@cliqz.com}
\and
\alignauthor
Josep M. Pujol\\
     \email{josep@cliqz.com}\\
     \affaddr{Cliqz GmbH}\\
     \affaddr{Arabellastraße 23}\\
     \affaddr{Munich, Germany}
}
\maketitle

\newcommand{\wtm}{\emph{WhoTracks.Me~}}

\begin{abstract}



Online tracking has become of increasing concern in recent years, however our
understanding of its extent to date has been limited to snapshots from web
crawls. Previous attempts to measure the tracking ecosystem, have been done using 
instrumented measurement platforms, which are not able to accurately capture 
how people interact with the web.
In this work we present a method for the measurement of tracking in the web
through a browser extension, as well as a method for the aggregation and collection
of this information which protects the privacy of participants.
We deployed this extension to more than 5 million users, enabling measurement across
multiple countries, ISPs and browser configurations, to give an accurate
picture of real-world tracking. 
The result is the largest and longest measurement of online tracking to date based 
on real users, covering 1.5 billion page loads gathered over 12 months. The data, 
detailing tracking behaviour over a year, is made publicly available
to help drive transparency around online tracking practices.

\end{abstract}




\section{Introduction}

On the modern web our actions are monitored on almost every page we visit by
third-party scripts which collect and aggregate data about users' activities 
and actions. A complex and dynamic ecosystem of advertising and
analytics has emerged to optimise the monetization of this data, and 
has grown to such an extent that 77\% of pages the average user will 
visit contain trackers~\cite{ghostrank_whitepaper2017}, and with 
individual trackers present on over 60\% of the top 1 million 
sites~\cite{DBLP:conf/ccs/EnglehardtN16}.

Monitoring this ecosystem has been the focus of recent efforts, looking into the methods 
used to fingerprint users and their devices~\cite{DBLP:conf/sp/NikiforakisKJKPV13}, 
and the extent to which these methods are being used across 
the web~\cite{DBLP:conf/ccs/AcarEEJND14}, and quantifying the value
exchanges taking place in online advertising~\cite{DBLP:conf/conext/CarrascosaMREL15,DBLP:conf/imc/PapadopoulosRKL17}.
There is a lack of transparency 
around which third-party services are present on pages, and what happens 
to the data they collect is a common concern. By monitoring this 
ecosystem we can drive awareness of the practices of these services, 
helping to inform users whether they are being tracked, and for 
what purpose. 
More transparency and consumer awareness of these practices can help drive
both consumer and regulatory pressure to change, and help researchers to better 
quantify the privacy and security implications caused by these services. 
With the EU's General Data Protection Regulation imminent at the time of writing, 
monitoring will be important to help detect violations.

Most previous work on measuring tracking prevalence at scale has focused 
on the engineering of crawlers which emulate a web browser visiting a series of pages~\cite{DBLP:conf/ccs/EnglehardtN16,DBLP:conf/eurosp/MerzdovnikHBNNS17}. 
These systems instrument the browser to collect detailed information 
about each page loaded. This method can scale well, however, bias is 
introduced by the choice of crawling platform, the physical location from which 
the crawl is run, and the sites chosen to be crawled. Further 
limitations exist around getting data from pages behind authentication walls, 
such as in online banking portals, e-commerce checkout pages, paywalled 
content, and `walled gardens' like Facebook and LinkedIn. Lastly, these
crawls capture an instantaneous state of the 
ecosystem, but do not enable longitudinal analysis. Longitudinal studies 
have typically been done on a smaller scale to one-off 
crawls~\cite{DBLP:conf/uss/LernerSKR16,DBLP:conf/www/KrishnamurthyW09}.

This work contributes a system for the continuous measurement of the presence 
of third-parties across the web, and the tracking methods employed.
This system gathers measurements via a large population of users 
who consent to data collection via a browser extension. We deploy a 
monitoring mechanism which collects data on third-party trackers for pages 
users visit, and employ a privacy-by-design methodology to ensure potential 
identifiable data or identifiers are removed on the client side before 
transmission. This enables measurement of tracking as observed by real 
users during normal browsing activities, at scale, across multiple browsers 
and physical locations, while respecting the privacy of the users collecting 
the data.
This overcomes may of the issues encountered by crawl-based analyses 
of tracking.

Our method,using instrumented browsers distributed to users who consent to gathering 
data during their normal browsing activity can achieve a greater scale 
than crawling. In previous work, we analysed 21 million pages loaded by 
200,000 users in Germany~\cite{DBLP:conf/www/YuMMP16}, and analysis of 
data collected from Ghostery's GhostRank covered 440 million pages for 
850,000 users~\cite{ghostrank_whitepaper2017}. In this paper we present 
the \wtm dataset, which contains aggregated data on third-party 
presence and tracking, released monthly. The data is generated by Ghostery 
and Cliqz users 
who have consented to anonymized HumanWeb~\cite{human_web} data collection. This generates 
data on an average of 100 million page loads per month, increasing to over 300
million since April 2018, and currently 
spans 12 months\footnote{May 2017 to April 2018}.


This paper is organised as follows. In Section~\ref{sec:measuring} we describe 
how online tracking can be measured at scale during normal browser usage. We 
also describe common tracking methods and how they can be detected using 
browser extension APIs. In Section~\ref{sec:privacy} we outline our approach to 
collection of the page load data, and how we prevent this data 
from being deanonymizable. Section~\ref{sec:dataagg} covers how we aggregate 
the collected data and generate meaningful statistics to describe the tracker 
ecosystem. We also describe our database which maps over 1000 tracker domains 
to services and companies which operate them. A selection of results are 
presented in Section~\ref{sec:results}, which show the extent of tracking 
which we have measured from 12 months of data, 
from a total of ~1.5 billion page loads.

The work makes the following contributions:
\begin{itemize}[noitemsep,topsep=0pt]
    
    \item The largest longitudinal study of online tracking to date, in terms 
    of number of pages and sites analysed, with a total of 1.5 billion pages 
    analysed, and data on around 950 trackers and 1300\footnote{We intend to increase these numbers as our database grows.} popular websites published under a permissive Creative
    Commons license.
    
    \item A public data set containing aggregated statistics on trackers 
    and websites across the web. 
    
    \item An open database to attribute common third-party domains 
    to services and companies, containing over 1000 tracker entries.
    
    \item A method and implementation of a system for measuring tracking 
    context in the browser, including fingerprinting detection based 
    on~\cite{DBLP:conf/www/YuMMP16}.
    
    \item A system for the collection of the measured page load data which 
    safeguards the privacy of the users from whom the data originates by 
    removing or obfuscating any potential identifiable information in individual messages, and 
    removing data which could be used to link messages together.
    
    \item A website providing information based on the collected data for interested
    users, and containing educational resources about online tracking.
    
    \item Results, reproducing findings of previous tracking studies, showing 
    trends in online tracking over the last year, and providing new insights on previously
    unmeasured tracking.

\end{itemize}

\section{Measuring Online Tracking}\label{sec:measuring}

Online tracking can be characterised as the collection of data about 
user interactions during the course of their web browsing. This can 
range from simply recording which types of browser access a particular 
page, to tracking all mouse movements and keystrokes. Of most concern to 
privacy researchers is the correlation and linkage of the data points 
from individual users across multiple web pages and web sites, primarily 
because of the privacy side-effects this entails: such histories, linked 
with identifiers, even when pseudo-anonymous, can be easily associated with 
individuals to whom they belong~\cite{DBLP:conf/www/SuSGN17}.

In this work we aim to measure the extent of this latter kind of tracking: 
the collection of linkable data points which generate a subset of users' 
browsing histories. As with other studies~\cite{DBLP:conf/ccs/EnglehardtN16,DBLP:conf/eurosp/MerzdovnikHBNNS17,DBLP:conf/www/KrishnamurthyW09,DBLP:conf/www/CahnABM16}, we do this by instrumenting the browser to 
observe the requests made from each page visited, and looking for 
evidence of identifiers which could be used to link messages together. 
Unlike other studies, which generally set up automated crawls to popular 
domains, we deploy our probe to users of the Cliqz and Ghostery browser 
extensions. This gives several advantages:

\begin{itemize}[noitemsep,topsep=0pt]

    \item \emph{Scale}: The probe is deployed to over 5 million users, 
    which gives us up to 350 million page load measurements per month. 
    Such scale cannot practically be achieved with crawlers.

    \item \emph{Client diversity}: With over 5 million users, we can 
    obtain measurements from a myriad of network and system environments. 
    This includes network location, ISP, Operating System, browser software 
    and version, browser extensions and third-party software. All of these 
    factors may have some influence on observed tracking. Previous studies 
    using crawling suffer from a monoculture imposed by tooling 
    limitations: Firefox on Linux in an Amazon data-centre.

    \item \emph{The non-public web}: Stateless web crawling limits one's 
    access to the public web only. These are pages which are accessible 
    without any login or user-interaction required. This excludes a significant 
    proportion of the web were tracking occurs, such as during payments on 
    E-commerce sites, when accessing online banking, or on `walled-gardens' such
    as Facebook~\cite{DBLP:conf/imc/KaizerG16}.

\end{itemize}

The downside of this approach is that when collecting data from real 
users as they browse the web, there could be privacy side-effects in the 
data collected. The aim is to be able to measure the extent of tracking, but 
without collecting anything which could identify individuals, or even having 
any data value that someone may consider private. Therefore, great care must 
be taken in the collection methodology: what data can and cannot be collected, 
and how to transmit this privately. Due to these constraints, the data we can 
collect is of much lower resolution as what can be collected from crawling. 
Therefore these two approaches can complement each other in this regard. 
We describe our methodology of privacy-preserving data collection in this paper.

\subsection{Tracking: a primer}

Tracking can be defined as collecting data points over multiple different 
web pages and sites, which can be linked to individual users via a unique 
user identifier. The generation of these identifiers can be \emph{stateful},
where the client browser saves an identifier locally which can be retrieved 
at a later time, or \emph{stateless}, where information about the browser 
and/or network is used to create a unique fingerprint. In this section 
we summarise the common usage of these methods:

\subsubsection{Stateful tracking}

Stateful tracking utilises mechanisms in protocol and browser APIs in order
to have the browser save an identifier of the tracking server's choosing, 
which can be retrieved and sent when a subsequent request is made to 
the same tracker.


The most common method is to utilise browser cookies.
As this mechanism is implemented by the browser, it is a client-side decision 
whether to honour this protocol, and how long to keep the cookies. Almost 
all browsers offer the option to block cookies for third-party domains when 
loading a web page, which would prevent this kind of tracking. However, 
browsers have defaulted to allow all cookies since the cookie specification was proposed, leading 
to many services and widgets (such as third-party payment and booking 
providers) relying on third-party cookies to function.

Other stateful methods include the JavaScript \\\texttt{localStorage} 
API \cite{MDN:LocalStorage}, which enables Javascript code to save data 
on the client side, and Cache-based methods using \texttt{E-Tags}~\cite{MDN:E-Tag}.

\subsubsection{Stateless tracking}

Stateless tracking combines information about the target system via 
browser APIs and network information, which, when combined, creates 
a unique and persistent identifier for this device or 
browser~\cite{DBLP:conf/pet/Eckersley10,DBLP:conf/sp/NikiforakisKJKPV13}. 
It differs from stateful methods in that this value is a product of the 
host system, rather than a saved state, and therefore cannot be deleted
or cleared by the user.

Certain hardware attributes, which on their own may not be unique, when combined create a unique digital fingerprint, which renders it possible to identify a particular browser on a particular device~\cite{DBLP:conf/pet/Eckersley10}. This method will 
usually require code execution, either via JavaScript or Flash, which is 
enable gather the data from APIs which provide device attributes like 
the device resolution, browser window size, installed fonts and plugins, 
etc~\cite{DBLP:conf/sp/NikiforakisKJKPV13}. More advanced methods leverage observations of the ways different 
hardware render HTML Canvas data~\cite{DBLP:conf/ccs/AcarEEJND14,bcookies:mowery} or manipulate audio data in order to 
generate fingerprints~\cite{DBLP:conf/ccs/EnglehardtN16}.

\subsubsection{Measuring Tracking Methods}

In most cases, both \emph{stateful} and \emph{stateless} tracking can be 
measured from the browser. Measurement of \emph{stateful} tracking is made 
easier by the origin requirements of the APIs being used. Both Cookies and 
localStorage sandbox data according to the domain name used 
by the accessing resource. For example, if a cookie is set for the 
domain \texttt{track.example.com}, this cookie can only be sent for requests 
to this address. This necessitates that trackers using these methods must
always use the same domain in order to track across different sites. Thus, 
this origin requirement enables us measure a particular tracker's 
presence across the web via the presence of a particular third-party 
domain---the identifier cannot be read by other domains

Stateless tracking does not have the same origin constraints as stateful 
tracking, therefore fingerprints could be transmitted to different domains, 
and then aggregated on the server side. Even though the use of stateful tracking is easier, due to the prevalence of browsers which will accept third-party cookies, we find that most trackers still centralise their endpoints. This is true also when 3rd parties engage in stateless tracking.

As stateless tracking uses legitimate browser APIs, we cannot assume 
simply that the use of these API implies that tracking is occurring. 
We use a method, based on our previous work, of detecting the transmission 
of data values which are unique to individual users~\cite{DBLP:conf/www/YuMMP16}. 
We detect on the client side which values are unique based on a 
$k$-anonymity constraint: values which have been seen by fewer than 
$k$ other users are considered as \emph{unsafe} with respect to privacy. 
We can use this method as a proxy to measure attempted transmission of fingerprints 
generated with stateless tracking, as well as attempts to transmit 
identifiers from stateful methods over different channels.

Note that these detection methods assume that trackers are not 
obfuscating the identifiers they generate.

\subsection{Browser Instrumentation}\label{sec:instrumentation}

We measure tracking in the browser using a browser extension. This 
enables us to observe all requests leaving the browser and 
determining if they are in a tracking context or not. For each page 
loaded by the user, we are able to build a graph of the third-party 
requests made and collect metadata for each.

HTTP and HTTPS requests leaving a browser can be observed using the 
\texttt{webRequest} API~\cite{webrequest_api}. 
This is a common API available on all major desktop web browsers. 
It provides hooks to listen to various stages of the lifecycle 
of a request, from \texttt{onBeforeRequest}, when the browser 
has initially created the intent to make a request, to 
\texttt{onCompleted}, once the entire request response has been received. 
These listeners receive metadata about the request at that point, 
including the url, resource type, tab from which the request 
originated, and request and response headers.

We first implement a system for aggregating information on a page 
load in the browser, enabling metadata, in the form of counters, 
to be added for each third-party domain contacted during the page 
load. We define a page load as being:

\begin{itemize}[noitemsep,topsep=0pt]

    \item Created with a web request of type \texttt{main\_frame} in a tab;

    \item Containing the \texttt{hostname} and \texttt{path} extracted
    from the \texttt{URL} of the main frame request;

    \item Ending when another web request of type \texttt{main\_frame} is 
    observed for the same tab, or the tab is closed.

\end{itemize}

For each subsequent request for this tab, we assess whether 
the hostname in the url is third-party or not. This is done by comparing 
the Top-Level-Domain$+1$ (TLD$+1$)\footnote{Top level domain plus first subdomain.}
forms of the page load hostname to that of 
the outgoing request. If they do not match, we add this domain as a 
third-party to the page load.

We collect metadata on third-party requests in three stages of the 
\texttt{webRequest} API: \texttt{onBeforeRequest}, \\\texttt{onBeforeSendHeaders},
\texttt{onHeadersReceived}.

In \texttt{onBeforeRequest} we first increment a counter to track 
    the number of requests made for this domain. Additionally we count:
        \begin{itemize}[noitemsep,topsep=0pt]
            \item the HTTP method of the request (GET or POST);
            \item if data is being carried in the url, for example in the query
            string or parameter string;
            \item the HTTP scheme (HTTP or HTTPS);
            \item whether the request comes from the main frame or a sub 
            frame of the page;
            \item the content type of the request (as provided by 
            the \texttt{webRequest} API);
            \item if any of the data in the url is a user identifier, according 
            to the algorithm from~\cite{DBLP:conf/www/YuMMP16};
        \end{itemize}
In \texttt{onBeforeSendHeaders} we are able to read information about 
    the headers the browser will send with this request, and can therefore 
    count whether cookies will be sent with this request.

In \texttt{onHeadersReceived} we see the response headers from 
    the server. We count:
        \begin{itemize}[noitemsep,topsep=0pt]
            \item that this handler was called, to be compared with 
            the \texttt{onBeforeRequest} count;
            \item the response code returned by the server;
            \item the content-length of the response (aggregated for all seen third-party requests);
            \item whether the response was served by the browser cache or not;
            \item whether a \texttt{Set-Cookie} header was sent by the server;
            \item the origin country of the responding server
            (based on a geoip lookup of the IP address\footnote{We use the MaxMind database for this purpose: \url{https://dev.maxmind.com/geoip/geoip2/geolite2/}}).
        \end{itemize}

As this code runs alongside Ghostery's blocking, we can also measure if requests were blocked by
this extension. Depending on user configuration, this may be category related blocking, specific
block rules, or based on Adblock blocklists.

Together, these signals give us a a high level overview of what 
third-parties are doing in each page load:

\begin{itemize}[noitemsep,topsep=0pt]
    
    \item Cookie's sent and \texttt{Set-Cookie} headers received 
    (in a third-party context) can indicate \emph{stateful} tracking via Cookies.
    Empirical evaluation shows that the use of non-tracking cookies by third-parties
    is limited.
    
    \item HTTP requests on HTTPS pages show third-parties causing 
    mixed-content warnings, and potentially leaking private information 
    over unencrypted channels.
    
    \item The context of requests (main or sub frames) indicate how 
    much access to the main document is given to the third-party.
    
    \item The content types of requests can tell us if the third-party 
    is permitted to load scripts, what type of content they are loading 
    (e.g. images or videos), and if they are using tracking APIs such 
    as beacons~\cite{w3c_beacon}.
    
    \item The presence of user identifiers tells us that the third-party 
    is transmitting fingerprints with requests, such as viewport sizes, 
    or other tracking parameters.
    
    \item The difference between the number of requests seen by 
    the \texttt{onBeforeRequest} and \texttt{onHeadersReceived} handlers 
    indicates the presence of external blocking of this third-party, either at the 
    network level or by another browser extension. We also measure if the
    extension hosting the measurement code blocked the request. This gives a
    measure of actual blocking due to Ghostery or Adblocker blocklists in the wild.

\end{itemize}

Once the described data on a page load has been collected, it is transmitted 
as a payload containing: the page's protocol (HTTP or HTTPS), the first-party 
hostname and path, and the set of third-parties on the page ($\textit{TP}$).

\begin{equation}
    \textit{pageload} = \langle \textit{protocol},\textit{hostname},\textit{path},\textit{TP} \rangle\\
\end{equation}

The set of third-parties simply contain the third-party hostnames 
with their associated counters:
\begin{equation}
    \textit{TP} = \{\langle \texttt{hostname}, C \rangle,\ldots\}
\end{equation}

The nature of this data already takes steps to avoid recording at a 
level of detail which could cause privacy side-effects. In 
Section~\ref{sec:privacy} we will describe these steps, and the further 
steps we take before transmitting this data, and in the transmission 
phase to prevent any linkage between any page load messages, nor any 
personal information in any individual message.


\section{Privacy-Preserving Data Collection}\label{sec:privacy}

The described instrumentation collects information and metadata 
about pages loaded during users' normal web browsing activities. 
The collection of this information creates two main privacy 
challenges: First, an individual page load message could contain 
information to identify the individual who visited this page, 
compromising their privacy. Second, should it be possible to 
group together a subset of page 
load messages from an individual user,
deanonymization becomes both easier, and of greater 
impact~\cite{DBLP:conf/www/SuSGN17,CCC/Eckert/2016}. In this 
section we discuss how these attacks could be 
exploited based-on the data we are collecting, and then, 
how we mitigate them.


\subsection{Preventing message deanonymisation}

The first attack attempts to find information in a \emph{pageload} 
message which can be linked to an individual or otherwise leak 
private information. We can enumerate some possible attack vectors:

\textbf{Attack 1}. The first-party \emph{hostname} may be private. Network 
    routers or DNS servers can arbitrarily create new hostnames 
    which may be used for private organisation pages. A page load 
    with such as hostname may then identify an individual's network 
    or organisation.

\textbf{Attack 2}. The \emph{hostname} \emph{path} combination often gives 
    access to private information, for example sharing links from
    services such as Dropbox, Google Drive and others would give
    access to the same resources if collected. Similarly password
    reset urls could give access to user accounts.

\textbf{Attack 3}. \emph{hostname} and \emph{path} combinations 
    which are access protected to specific individuals could leak their 
    identity if collected. For example, the twitter analytics 
    page \url{https://analytics.twitter.com/user/jack/home} can
    only be visited by the user with twitter handle \texttt{jack}~\cite{modi/green_tracker}.

\textbf{Attack 4}. Third-party hostnames may contain user identifying information. 
    For example, if an API call is made containing a user identifier
    in the hostname, it could be exploited to discover more about 
    the user. While this is bad practice, as the user identifier is
    then leaked even for HTTPS connections, we have observed this 
    in the wild~\cite{microsoft_domains}.

We mitigate attacks 1. and 2. by only transmitting a truncated MD5 hash\footnote{While using truncated
hashes does not bring improved privacy properties, it does provide
plausible deniability about values in the data.}
of the first-party \emph{hostname} and \emph{path} fields. By 
obfuscating the actual values of these fields we are still able 
to reason about popular websites and pages --- the hashes of public 
pages can be looked up using a reverse dictionary attack --- but private domains 
would be difficult to brute force, and private paths (e.g. password 
reset or document sharing links) are unfeasible. 
Therefore this treatment has desirable privacy properties, 
allowing us to still collect information about private pages without 
compromising their privacy and that of their users.

This treatment also mitigates some variants of attack 3., however 
for sites with a predictable url structure and public usernames 
(like in our twitter analytics example), it remains possible to 
lookup specific users by reconstructing their personal private 
url. We prevent this by further truncating the path before hashing 
to just the first level path, i.e. \texttt{/user/jack/home} would 
be truncated to \texttt{/user/} before hashing.

Attack 4. cannot be mitigated with the hashing technique, as we need to 
collect third-party domains in order to discover new trackers. We can, 
however, detect domains possibly using unique identifiers by counting 
the cardinality of subdomains for a particular domain, as well as 
checking that these domains persist over time. After manually 
checking that user identifiers are sent for this domain, we push 
a rule to clients which will remove the user identifier portion of 
these hostnames. We also report these cases to the service providers, 
as this practice represents a privacy leak to bad actors on the 
network. We can further reduce the probability of collecting unique 
subdomains by truncating all domains to TLD$+2$ level.


\subsection{Preventing message linkage}

Even if individual messages cannot be deanonymised, if messages can be 
linked it is possible that as a group they can be deanonymised, as shown
in recent examples deanonymising public datasets~\cite{DBLP:conf/www/SuSGN17,CCC/Eckert/2016}. Furthermore, 
if an individual message happens to leak a small amount 
of information, once linked with others the privacy compromise becomes 
much greater. Therefore, we aim to prevent any two \emph{pageload} 
messages from being linkable to one-another.

The linking of messages requires the message sent from an individual user to be both unique, so that it does not intersect with others', and persistent, so that it can be used to link multiple messages 
together. We can enumerate some possible attacks:

Referring to attack 4 from the previous section may also be used for linkage, 
    if the unique hostname is used over several popular sites. For 
    example a case we found with Microsoft accounts was prevalent across 
    all Microsoft's web properties when a user was logged in. The third-party 
    domain was specific to their account and did not change over time. 
    This third-party domain would therefore be used to link all visits to 
    Microsoft sites indefinitely.

\textbf{Attack 5}. In a previous version of our browser instrumentation we collected 
    the paths of third-party resources as truncated hashes. However, some
    resource paths could then be used for message linkage, for example 
    avatars from third-party services such as Gravatar could be used to 
    link visits on sites which display this avatar on every page for the 
    logged in user. For this reason we removed collection of these paths.

\textbf{Attack 6}. Some third-party requests can be injected into pages by other
    entities between the web and the user. ISPs can intercept insecure web 
    traffic, Anti-virus software often stands as a Man in the Middle to all 
    connections from the browser, and browser extensions can also inject 
    content in the page via Content scripts. Any of these entities can cause 
    additional third-parties to appear on page loads. It is possible that a 
    combination of injected third-parties could become unique enough to act 
    as a fingerprint of the user which could link page loads together.

\textbf{Attack 7}. When data is uploaded from clients to our servers we could log 
    the originating IP addresses of the senders in order to group the 
    messages together, or utilise a stateful method to transmit user 
    identifiers with the data.

We have already presented mitigations for the first two attacks. Attack 6. 
is difficult to mitigate for two reasons. Firstly, of the injected 
third-parties which we do detect, we cannot quantify the number of 
distinct users affected from the data that we collect. Therefore, 
it is not possible at the moment to calculate if certain combinations of 
third-parties would be able to uniquely identify an individual user. 
Secondly, a large proportion of these third-parties are injected 
by malware or other malicious actors, which implies an unstable ecosystem,
where, as extensions get blocked 
and domains get seized, the set of injected third-parties will 
change. This also will have the effect that the persistence of 
the links will be limited. Despite this we aim to develop a 
mitigation method as part of our future work.

Attack 7 looks at the case where we ourselves might be either 
malicious or negligent as the data collector, creating a log
which could be used to link the collected page loads back to 
pseudo-anonymous identifiers. It is important, that when monitoring
trackers, we do not unintentionally become one ourselves. 
Trust is required, both that our 
client side code does not generate identifiers to be transmitted 
to the server along side the data, and that the server does not
log IP addresses from which messages are received.

Trust in the client side is achieved by having the extension 
code open-sourced\footnote{https://github.com/cliqz-oss/browser-core}, 
and the extension store review and distribution processes should, in 
theory, prevent a malicious patch being pushed to diverge from the 
public code. Furthermore, extensions can be audited in the browser 
to allow independent inspection of requests leaving the browser.

In order to allow the client to trust that the server is not using
network fingerprints to link messages, we have developed a system 
whereby data is transmitted via proxies that can be operated by independent 
entities. Encryption is employed such that these proxies cannot 
read or infer anything about that transmitted data. The scheme 
is therefore configured such that the data collection server only 
sees data messages---striped of user IPs---coming from the 
proxies. The proxies see user IP addresses  and encrypted blobs 
of data. Proxies visibility of message transmissions is limited by 
load-balancing, which partitions the message space between the 
acting proxies, limiting how much metadata each is able to 
collect. The client-side part of this system also implements 
message delay and re-ordering to prevent timing-based correlations~\cite{human_web}.

The deployment of this system means that, if the user trusts the 
client-side implementation of this protocol, and the independence 
of the proxies, then he does not have to trust our data collection 
server to be sure we are not able to link messages together.

\subsection{Privacy Evaluation}

We have evaluated the risks in collecting the data gathered through 
our described browser instrumentation, and several steps which we 
take to mitigate and prevent these risks from being exploitable. 
We cannot prove completely anonymized data 
collection - we have made several improvements in response to 
findings from both internal and independent external audits of 
this data - however we regard this methodology as being 
robust, and if the data were to be leaked we are confident that 
the privacy consequences would be minimal.


\section{Data aggregation}\label{sec:dataagg}

In this section we describe how the collected page load messages 
are aggregated to provide high-level statistics which describe 
the tracking ecosystem. 

In previous studies of the tracking ecosystem, third-party domains have been 
truncated to TLD+1 level, and then aggregated. The reach of, for 
example \hfill\break\url{google-analytics.com}, will be then reported as the 
number of sites which have this domain as a third-party. This is a 
simple and easily understandable aggregation method, however it has 
some shortcomings:
\begin{itemize}[noitemsep,topsep=0pt]

    \item A domain name is not always transparent. For example it 
    will not be apparent to everyone that the domain \texttt{2mdn.net} 
    is operated by Google's Doubleclick advertising network. It is 
    important that the entities of the aggregation are meaningful and 
    transparent.

    \item Domain level aggregation will duplicate information for service 
    which use multiple domains in parallel. For example Facebook uses 
    \texttt{facebook.net} to serve their tracking script, and then send 
    tracking pixel requests to \texttt{facebook.com}, where the Facebook 
    tracking cookie resides. According to domain semantics these are 
    separately registered domains, though they will always occur together 
    on web pages. Therefore reporting these two domains separately is redundant, 
    and potentially misleading, as one might assume that the reach of 
    the two entities can be added, when in fact they intersect almost 
    entirely.

    \item Domain level aggregation will hide tracker entities who use a 
    service on a subdomain owned by another organisation. The prime case 
    here is Amazon's \texttt{cloudfront.com} CDN service. Several trackers 
    simply use the randomly assigned \texttt{cloudfront.com} domains rather 
    than use a CNAME to point to their own domain. For example 
    New Relic\footnote{New Relic is an performance analytics service which reaches over 4\% of web traffic as measured by our data} 
    sometimes uses the \texttt{d1ros97qkrwjf5.cloudfront.net} domain. 
    If we aggregate all Cloudfront domains together, the information 
    about different trackers is lost.
\end{itemize}

We solve these issues by using a manually curated database, based on
Ghostery's~\cite{ghostery} tracker database, which maps domains and subdomains 
to the services and/or companies they are know to operate under, as a base. 
For a given domain, the database may contain multiple subdomains 
at different levels which are mapped to different services. When 
aggregating domains, we then find the matching $TLD+N$ domain in 
the database, with maximal $N$. i.e. if we have mappings for 
\texttt{a.example.com}, \texttt{b.example.com} and \texttt{example.com}, 
then \texttt{a.a.example.com} would match to \texttt{a.example.com}, 
while \texttt{c.example.com} would be caught by the catch-all 
\texttt{example.com} mapping. These mappings allow us to split and 
aggregate domains in order to best describe different 
tracking entities.




\subsection{Different measurements of reach}

The page load data we collect allows us to measure tracker and 
companies' reach in different ways. We define a tracker or 
company's `reach' as the proportion of the web in which they are 
included as a third-party. This is done by counting the number 
of distinct page loads where the tracker occurs: 

\begin{equation}
    \textit{reach} = \frac{|\textit{page loads including tracker}|}{|\textit{page loads}|}
    \label{eq:reach}
\end{equation}

Alternatively, we can measure `site reach', which is the proportion of 
websites (unique first-party hostnames) on which this tracker has 
been seen at least once.

\begin{equation}
    \textit{site reach} = \frac{|\textit{unique websites where tracker was seen}|}{|\textit{unique websites}|}
    \label{eq:site-reach}
\end{equation}

Differences between these metrics are instructive: \emph{reach} is 
weighted implicitly by site popularity---a high \emph{reach} 
combined with low \emph{site reach} indicates a service which is 
primarily on popular sites, and is loaded a high proportion of the 
time on these sites. The inverse relation---low \emph{reach} and 
high \emph{site reach}---could be a tracker common on low traffic 
sites, or one which has the ability to be loaded on many sites 
(for example via high reach advertising networks), however does so rarely.

\subsection{Aggregation of instrumentation counters}

The reach metrics described are based on presence---when requests 
occur in a page to specific third parties. In 
Section~\ref{sec:instrumentation} we described other counters we collect 
in order to measure use of potential tracking vectors. We aggregate these 
statistics by counting the number of pages where these methods are invoked 
at least once during the page load, then report this metric as the 
proportion of the tracker's \emph{reach} which used this method. 
We report:

\begin{itemize}[noitemsep,topsep=0pt]
    \item Cookie tracking context -- Cookies sent with request, or 
    server responded with a \texttt{Set-Cookie} header.
    \item Fingerprinting context -- User identifier detected in the 
    request (as per~\cite{DBLP:conf/www/YuMMP16}).
    \item Tracking context -- Either cookie tracking or fingerprinting context, inclusive.
    \item Secure context -- Only HTTPS requests for the page load.
    \item Content types -- Pages where specific resource types were loaded by 
    the tracker (e.g. scripts, iframes, plugins)
    \item Blocking effect -- How often the tracker is affected by blocklist-based blockers.
\end{itemize}

Furthermore we report the mean number of third-party requests per page for each tracker, 
and the subset of these requests in a tracking context. 


\section{Results}\label{sec:results}




Most studies analysing the tracking landscape have generally been 
performed in the context of one off measurements \cite{DBLP:conf/ccs/EnglehardtN16} or 
longitudinal surveys with limited scale and scope 
\cite{DBLP:conf/www/KrishnamurthyW09, DBLP:conf/uss/LernerSKR16}. 
In the remainder of this section, we look at these two
perspectives: dissecting the tracking landscape data at a snapshot in time, 
and analysing longitudinal trends that reveal trends and could inform policy.

We structure each subsection in a way that describes measurements in the perspective 
of the parties involved: websites, third parties and users. This enables us to better 
measure the dynamics of the industry.

It is important to note that unlike other studies, in which the
measurement platform does not interact with websites in the same way
real users would, \cite{DBLP:conf/ccs/EnglehardtN16}, the data which will
be subject to our analysis, has been generated by users of our browser
extension over the course of the last year. As such, the behaviour of 
trackers and websites is what we see in reality.

The data spans from May 2017 to April 2018, amounting to a total number of page 
loads of 1.5 billion. This is the largest dataset on web tracking to our 
knowledge \cite{DBLP:conf/ccs/EnglehardtN16}.

\subsection{Snapshot in Time}

We will be looking at the data from April 2018, composed of roughly
340 million page loads, and filtering the top 1330 most visited websites. We
measure that 71\% of the traffic to these sites contains tracking. The average
number of trackers per site is 8, and the average number of tracking requests per 
page load 17.

\subsubsection{First parties}

\begin{figure}[ht]
  \includegraphics[width=\linewidth]{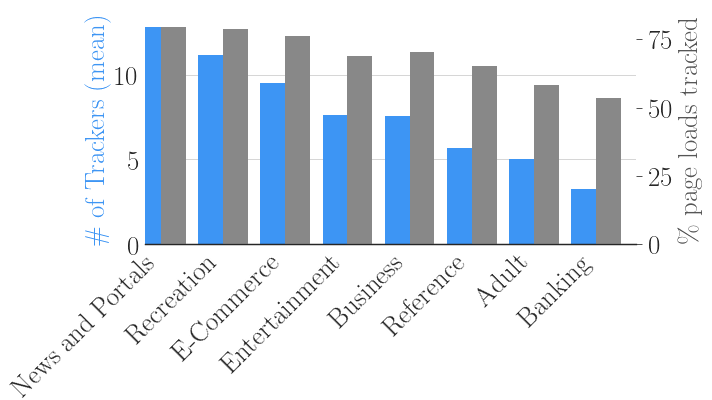}
  \caption{Tracking by website categories}
  \label{fig:snap-website-type-trackers}
\end{figure}

In Figure~\ref{fig:snap-website-type-trackers} we see that websites in the category of 
News and Portals have the highest number of  third parties  at approximately 
13 per page on average, with tracking occurring on 79 \% of the measured page 
loads. Banking websites tend to have the lowest number of third parties as well 
as a lower percentage of page loads where tracking occurs.

\subsubsection{The most prevalent third parties}
Third parties often provide functionality that is not immediately 
distinguishable from or visible in the website they are present on. Hence, 
to achieve transparency and understand the tracking market structure, 
estimating the prevalence of a particular tracker defined in terms of the 
fraction of web traffic they are present on (reach), is important.

\begin{figure}[ht]
  \includegraphics[width=\linewidth]{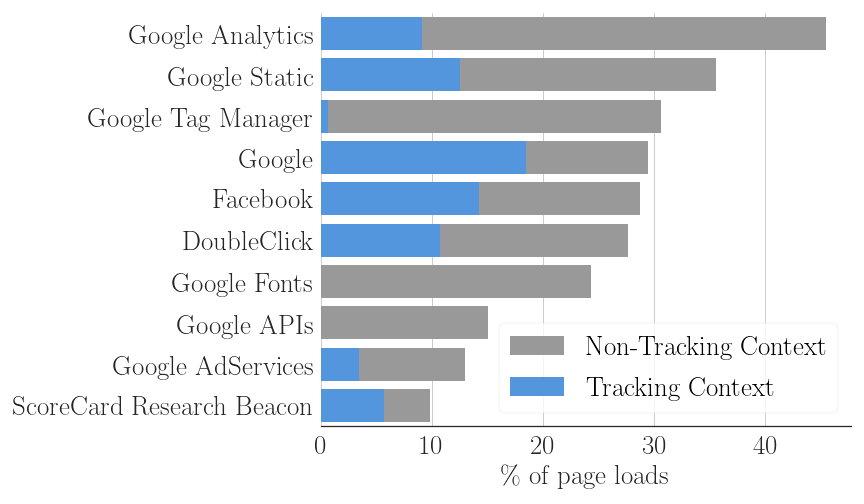}
  \caption{Top 10 third parties by reach}
  \label{fig:snap-tracker-reach}
\end{figure}

If we look at the top 10 third parties in Figure~\ref{fig:snap-tracker-reach}, we see that
Google Analytics has the highest reach, being present on roughly 
46\% of the measured web traffic, and 8 out of the top 10 third parties
are operated by Google.

Note that third parties do not always operate in a tracking context, which 
given our definition of third-party tracking, means they do not always 
send unique user identifiers. For instance, Google APIs is mostly used 
to load other 3rd parties such as Google Fonts and other static scripts, 
which is why we see it largely operating in a non-tracking context.

\subsubsection{From trackers to organisations}
By clustering third parties under parent organisations, we can also measure 
the reach of the latter.

\begin{figure}[ht]
  \includegraphics[width=\linewidth]{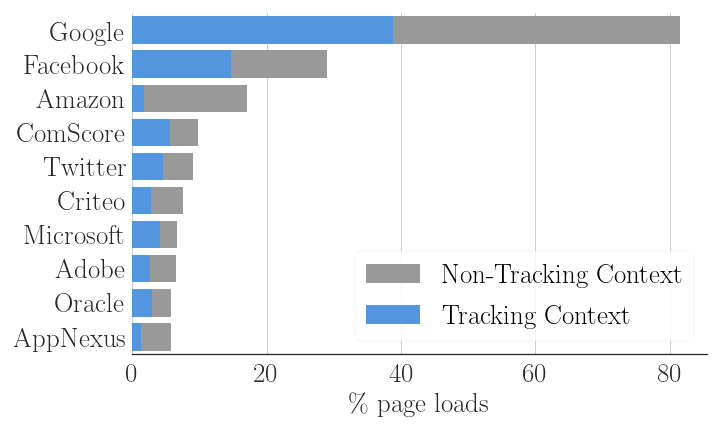}
  \caption{Top 10 organisations by reach}
  \label{fig:snap-company-reach}
\end{figure}

We observe that third-party scripts owned by Google are present in about 
82\% of the measured web traffic, and operate in a tracking context for slightly 
less than half that time. Facebook and Amazon follow next, and generally the
distribution of reach by organisation in Figure~\ref{fig:snap-company-reach} has
a long tail.

\subsubsection{Third Parties: categories and consequences}
\begin{figure}[ht]
  \includegraphics[width=\linewidth]{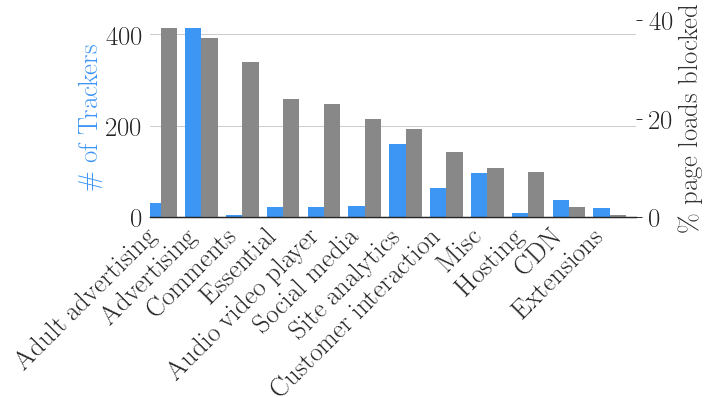}
  \caption{Third-party counts and block-rates by category}
  \label{fig:snap-trackers-by-category-blocked}
\end{figure}

Most third parties are loaded to perform certain functionality that websites need. 
Note how among third parties with the highest reach in Figure~\ref{fig:snap-trackers-by-category-blocked}, 
those that provide advertising services are predominant (left y-axis in blue), 
representing almost half of the trackers analysed in this study. In the same
figure, we see the proportion of page loads containing a tracker of a given category was
blocked by an ad-blocker.

Note, that as our reach measurement occurs before blocking, these block rates
are not reflected on the third-party and company reach we have already reported.

\subsubsection{Reach to Site Reach ratio}
Besides web traffic presence, we can also measure the first 
party presence for these third parties (site reach). The ratio of reach to 
site reach tells us an interesting story about the nature of the third party. 
The higher this ratio is, the more it  suggests the third party being popular 
on few popular domains, and  the lower it is, the more likely the third party 
could be some form of malicious software.

Take the example of DoublePimp with a a reach to site reach ratio of 28.8 
(reach: 0.8\% and site reach: $0.0002$\%), typically present on 
adult sites, and  particularly in a few popular ones.

Similarly, \texttt{eluxer.net}, with a reach to site reach ratio of 0.1, is a 
malicious extension which does insert tracking requests into pages as the
user browsers.

\subsubsection{A new breed of tracker}

Our data also measures a previously unmeasured type of tracker - those placed not by website owners
or ad networks, but by men in the middle. These are trackers which insert extra requests into pages
either by intercepting network traffic on a device, or using browser extensions. The largest of 
these trackers is the anti-virus vendor Kaspersky, whose software installs new root certificates on
the user's system in order to man-in-the-middle all requests from the operating system, and insert
tracking requests into every HTML document. This method enables the tracking of 2\% of total web
browsing (i.e. participants with this software installed represent 2\% of the collected page loads).

\begin{table}
    \centering
    \begin{tabular}{l|c|r}
    \textbf{Domain} & \textbf{Method} & \textbf{Reach}\\
    \hline
    kaspersky-labs.com & HTTP MITM & 2.0\% \\
    worldnaturenet.xyz & Malicious extension & 0.27\% \\
    eluxer.net & Malicious extension & 0.20\% \\
    ciuvo.com & Price comparison ext & 0.16\% \\
    comprigo.com & Price comparison ext & 0.15\% \\
    \end{tabular}
    \caption{Man in the middle (MITM) trackers}
    \label{tab:mitm_trackers}
\end{table}

Table~\ref{tab:mitm_trackers} shows the top 5 such trackers. From our investigations,
\texttt{worldnaturenet.xyz} and \texttt{eluxer.net} both appear to be extensions installed 
via malware, which then track and inject
advertising into pages. We were not able to determine the owners of these operations, but there are
several others with similar characteristics in our data. In contrast, the \texttt{ciuvo.com} and
\texttt{comprigo.com} browser extensions can be easily found, and the companies operating them.

\subsubsection{Regional Data flows}

In Section~\ref{sec:instrumentation} we noted that we can observe the IP address of the responding server, and from that use a GeoIP database to retrieve the country this server is situated in. Using this data, we can assess data flows from users in specific countries to trackers located in others. 
Table~\ref{tab:data_flow_mat} shows where third-party requests are loaded from for pages loaded from Australia, 
Germany, France, the UK, the Netherlands, Russia and the USA. 

We can see that in most cases the majority of page loads are tracked by servers located in the USA. Tracking of US users rarely goes abroad - 7\% of tracked pages make requests to Ireland - while in other regions US servers track on most pages. One exception is Russia, where Russian servers track marginally more pages than those based in the USA (64\% to 62\%).

\begin{table}
    \centering
    \begin{tabular}{l||r|r|r|r|r|r|r|r}
        \multirow{2}{*}{\textbf{From}} & \multicolumn{8}{c}{\textbf{\% pages with 3rd party request to}} \\
        & AU & DE & FR & GB & IE & NL & RU & US \\
        \hline \hline
        \textbf{AU} & 26 & 1 & 0 & 0 & 5 & 1 & 2 & \textbf{92} \\
        \textbf{DE} & 0 & 41 & 14 & 8 & 29 & 34 & 5 & \textbf{79} \\
        \textbf{FR} & 0 & 11 & 31 & 7 & 21 & 19 & 4 & \textbf{82} \\
        \textbf{GB} & 0 & 4 & 3 & 24 & 22 & 30 & 3 & \textbf{81} \\
        \textbf{NL} & 0 & 7 & 4 & 4 & 29 & 38 & 4 & \textbf{79} \\
        \textbf{RU} & 0 & 9 & 5 & 1 & 20 & 13 & \textbf{64} & 62 \\
        \textbf{US} & 0 & 1 & 1 & 1 & 7 & 2 & 2 & \textbf{98} \\
    \end{tabular}
    \caption{Locations of third-party services accessed for users in different geographical regions. }
    \label{tab:data_flow_mat}
\end{table}

Note, a limitation of this result is the validity of GeoIP results from some 
data centres. Notably, Google IPs always resolve to be located in the USA with the database 
we use, despite servers actually being located worldwide.

\subsection{Longitudinal}
Longitudinal studies have typically been done on a smaller scale 
to one-off crawls \cite{DBLP:conf/www/KrishnamurthyW09, DBLP:conf/uss/LernerSKR16}. 
Having a clear snapshot view of tracking at scale is important, but this 
often means the dynamics of tracking over time, are lost.

In this section, we explore the data at different levels of granularity from
measuring the data cost imposed on users by third parties to technology trends in
the tracking landscape.

\subsubsection{HTTPS Adoption}

Previous studies have highlighted the issue of insecure third-party calls
compromising the security and privacy of page
loads~\cite{DBLP:conf/ccs/EnglehardtN16}. In this work
we measure the protocol of outgoing third-party requests from the browser. 
We can use this measurement to detect the adoption rates of HTTPS across sites, 
and specifically which trackers are lagging behind on this metric.

\begin{table}
    \centering
    \begin{tabular}{l||r|r|r}
        \textbf{Entity} & \multicolumn{3}{c}{\textbf{All 3rd party requests secure}} \\
        & May 2017 & April 2018 & Change \\
        \hline \hline
        Top sites & 56.7\% & 81.1\% & $+24.4$\% \\
        News sites & 27.0\% & 68.0\% & $+41.0$\% \\
        \hline \hline
        Google Analytics & 64.0\% & 84.2\% & $+20.3$\% \\
        Facebook & 72.2\% & 83.9\% & $+11.7$\% \\
        AppNexus & 56.5\% & 83.5\% & $+27.0$\% \\
        Snowplow & 72.6\% & 46.3\% & $-26.4$\% \\
    \end{tabular}
    \caption{HTTPS Adoption of sites and trackers from May 2017 to April 2018}
    \label{tab:https}
\end{table}

Table~\ref{tab:https} shows how a selection of entities HTTPS usage has changed over the study period. We can see a strong increase in sites which have all third-party content loaded over HTTPS, from 57\% to 81\%. Certain categories of site lag behind in this regard though, namely news sites.

Looking at specific trackers, we can see dominant players such as Google Analytics and AppNexus successfully migrating clients to HTTPS over the year. Others, like Facebook, have had slower progress on this front.

In general, trackers improved their HTTPS usage over this period, with only 65 trackers (of 587 with data at both time points) not increasing HTTPS coverage. A small number of outliers showed a negative trend, for example Snowplow\footnote{\url{https://snowplowanalytics.com/}}, an analytics provider present on major news websites, including Le Monde and New York Times.


\subsubsection{Cost imposed on users}
As users navigate the web, they load content from websites they visit 
as well as the third parties present on the website. On average, 
for each first party page load, there are 17 third-party tracking requests. 
So beyond the privacy erosion, there is a material cost involved in this 
transaction. Previous studies have found that each extra third party added 
to the site will contribute to an increase of 2.5\% in the site's loading 
time \cite{ghostery/tracker_tax}. Here we measure the amount of data needed to 
load third-party content.


We take the sum of the \texttt{Content-Length} of all third-party requests in the 
top 1330 websites over the last year, and measure the variation in this data 
consumption overt time. The median content length per site from third parties
was 0.42MB with an interquartile range (IQR) of 0.18-1.5MB, down from 0.58MB 
(IQR 0.24-1.8MB) a year earlier.  The distribution has a long tail due to third 
parties offering Audio Video player services being part of the data.


\subsubsection{Tracking technologies over time}
There are several observations in how different content types are used in the
context of tracking. The following are measured:

\begin{itemize}[noitemsep,topsep=0pt]
    \item script: Javascript code (via a \texttt{<script>} tag or web worker).
    \item iframe: A subdocument (via \texttt{<frame>} or \texttt{<iframe>} elements).
    \item beacon: Requests sent through the Beacon API.
    \item image: \texttt{Image} and \texttt{imageset} resources.
    \item stylesheet: CSS files.
    \item font: Custom fonts.
    \item xhr: Requests made from scripts via the \texttt{XMLHttpRequest} or \texttt{fetch} APIs.
    \item plugin: Requests of \texttt{object} or \texttt{object\_subrequest} types, which are typically associated with browser plugins such as Flash.
    \item media: Requests loaded via \texttt{<video>} or \texttt{<audio>} HTML elements.
\end{itemize}

With this data we can see that, for example, during April 2018 Google Analytics loaded their script on each page load (97\% of the time), then registered the visit via an image (pixel) on 50\% of page loads. We also see that on 2.6\% of pages a request is also made via the Beacon API.

\begin{figure}[ht]
  \includegraphics[width=\linewidth]{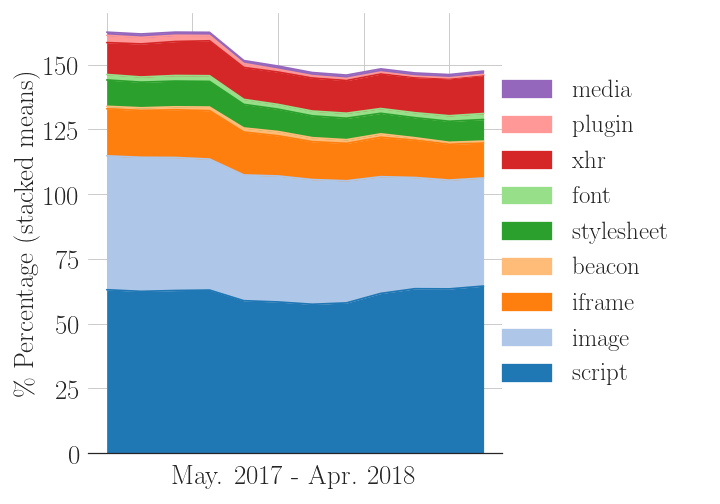}
  \caption{Content type usage for third parties}
  \label{fig:long-content-type}
\end{figure}

In Figure~\ref{fig:long-content-type} we see that scripts and images are the most
popular content types for tracking. Interestingly, beacons, originally designed to satisfy tracking use cases are encountered increasingly less.

\subsubsection{Reach by type of third party over time}
\begin{figure}[ht]
  \includegraphics[width=\linewidth]{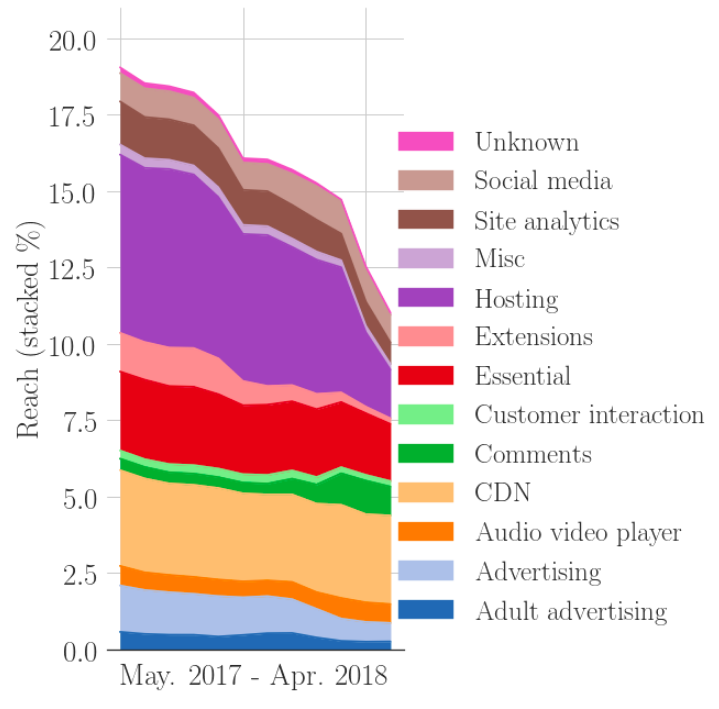}
  \caption{Reach over time by type of third party}
  \label{fig:long-tracker-type-mean-reach-over-time}
\end{figure}

The data also enables us to monitor the reach of third parties
over time. If grouped and averaged as in Figure~\ref{fig:long-tracker-type-mean-reach-over-time}
we observe almost an across-the-board decrease in the reach of third parties, most notably in the category of extensions that engage in MITM tracking. One explanation could be attributed to an increased adoption of ad-blockers.

\begin{figure}[ht]
  \includegraphics[width=\linewidth]{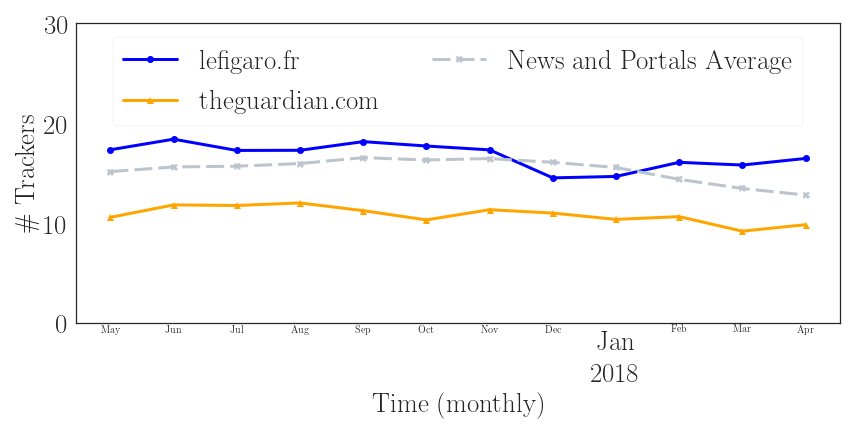}
  \caption{Third parties: The Guardian and Le Figaro}
  \label{fig:long-site-trackers-comparison-over-time}
\end{figure}

This analyses can be conducted at a more fine granular level, by monitoring 
the change in the average number of third parties in any given site. In 
Figure~\ref{fig:long-site-trackers-comparison-over-time} we compare the 
average number of third parties present on The Guardian and Le Figaro with the
industry average over the last year.

\subsection{Discussion}

Our results re-affirm previous findings: That significant numbers of third-parties are
loaded on each page a user visits across the web. The number of third-parties is the highest on
news websites, and where tracking is utilised. The number of trackers per page on a website
generally trends with the presence of advertising networks and the Adtech supply chain which
permits multiple parties to a bid to execute scripts on a page.

One surprising aspect may be the prevalence of tracking on business websites. This is again
tied to Adtech conversion measurement: business who advertise elsewhere on the web are
encouraged to install their ad vendor's scripts on their pages in order to attribute landings
to users who view particular ads. This enables fine-grained measurement of specific campaigns
across ad platforms.

Figure~\ref{fig:snap-trackers-by-category-blocked} confirms that the largest category of
third-parties is in advertising, but these are also the most heavily affected by blocking,
with almost 40\% of page loads seen by advertising trackers affected by blocking. This
provides a extra layer of nuance of previous reports how the level of ad-blocker adoption~\cite{pagefair},
showing the amount of blocking within a population of ad-blocker 
users\footnote{Ghostery and Cliqz both integrate an ad-blocker}, taking whitelisting 
and gaps in blocklists into account.

Our longitudinal analyses show a decline in the number of third-parties loaded on pages. We
may then infer that website owners are reducing the number of third-parties they allow to be
loaded on their pages. It could also be tied to changes in Adtech triggered by the GDPR, where
supply chains are being condensed in an attempt to become compliant, and to increase the 
chance of getting user consent for tracking~\cite{google_gdpr_consent}. However, one likely
larger contributor to this drop is the aforementioned ad-blocking. As well as the blocking we
measure from the resident browser extension, many users will have additional adblocking 
extensions or firewall rules installed to block certain third-parties. A side-effect of blocking
ad networks, this a lower reported number of third-parties on the page, as by blocking the
initial ad network request, subsequent vendors which would have been loaded by this first script
are not seen. This has the effect of reducing the number of third-parties measured.

Of note, and concerning for websites trying to become compliant with data protection law, is our
analysis of third-party content types. We measure that most of third-parties are permitted to load
scripts into publishers' pages, and this is the most common way in which third-party content is 
embedded.

This is firstly a security issue - scripts loaded in the main document of a website
have access to all page content, and can perform any action they wish. The prevalence of this
practice makes malvertising---the serving of malware via advertising networks---possible, and
presents a much larger attack surface against the site. In a recent when a third-party script
was compromised and started loading a cryptocurrency mining script in the website
of the Internet Commissioner's Office \texttt{ico.org.uk} in the UK and more than 
4000 other websites where this third party was present~\cite{helme_cryptojacking}.

Secondly, this is a compliance challenge. As scripts provide the third-parties with significant
capabilities to ex-filtrate data from the website in which they are embedded, to be
compliant website owners should require contracts to state the constraints under which the
third-party must operate, such that any consent that the first-party obtains for data
processing is valid for what the third-party actually does. Our position is that this is
likely overly burdensome, and the adoption of privacy-by-design solutions would be preferable,
where the system design enforces constraints on third-parties, and non-compliance is not
technically possible.

A positive result of our longitudinal analysis is the continuing adoption of HTTPS by both first
and third parties. A combination of nudges have encouraged providers to switch, making
certificates easier to obtain via services such as LetsEncrypt\footnote{\url{https://letsencrypt.org/}}, 
increased pressure from browser
vendors, blocking some kinds of mixed-content and UI changes such as showing warnings on forms on
insecure pages, and increased concerns about network eavesdroppers, such as ISPs. Progress, however,
is still dependant on the third-party vendors used, as our results show. Some services have
achieved better progress than others in this regard.

Note that our results for HTTPS adoption may over estimate in some aspects. A proportion of participants
(those using the Cliqz browser) have the HTTPS Everywhere\footnote{\url{https://www.eff.org/https-everywhere}}
installed and enabled by default, and this extension will prevent loading of insecure sites when a secure
version is available, thus increasing the reported HTTPS adoption rate.

Our results also measure a new kind of tracking - that of browser extensions, malware and other software
injecting requests into pages browsed by users. While the presence of spyware and malware in browser
extension stores is not new, our results provide a first look at its prevalence in the wild. We hope
that this data can be used by browser vendors to detect malicious extensions, or when users' privacy
could be compromised by malware on their system.



\section{WhoTracks.Me website}

One of the contributions is \wtm, a website that hosts the largest 
dataset of tracking on the web and detailed analysis of the growing body of data,
neatly organised around detailed profiles of first and third parties, respectively
referred to as websites and trackers.

For each website, we provide a list of data that infers the tracking 
landscape in that website. The per site third-party data includes, but is not limited to: 
the number of third parties detected to be present at an average page load of that
website as well as the total number of third parties observed in the last month;
the frequency of appearance for each third party;
the tracking mechanisms used by third parties in the site;
a distribution of services the present third parties perform on that page
Heavy use of data visualisations is made to make the data accessible to as wide 
a spectrum of an audience as possible.


Given the often obscure profiles of trackers, for each tracker we try to 
identify the organisation operating it and make the information accessible.
For each tracker profile, we provide the information needed to identify 
them; the list of domains it uses to collect data; the organisation that 
operates them; reach and site reach as defined in equations ~\ref{eq:reach} 
and ~\ref{eq:site-reach}), as well as the methods they use for tracking. 
Furthermore, we provide information on the distribution of the types of 
websites they are seen to be present, similar third parties and a list of sites
where it has been seen to be present. For an example, 
please visit a tracker profile on \wtm.




\subsection{Who is WhoTracks.Me for?}
\wtm is a monitoring and transparency tool. We have open
sourced data from more than 1.5 billion page loads per month, and 
plan to continue the effort. As tersely demonstrated in 
Section \ref{sec:results}, the possibilities for using 
the data are numerous and the users diverse:

\begin{itemize}[noitemsep,topsep=0pt]
    \item \textbf{Researchers} - Can use the open data to investigate
    tracking technologies, develop more comprehensive protection mechanisms 
    and threat models, investigate the underlying structure of online
    tracking as a marketplace etc.
    
    \item \textbf{Regulators} - The ability to access both detailed 
    snapshots of tracking data as well as observe entities over time, 
    enables regulators to use \wtm as a monitoring tool to 
    measure the effect of regulations like the General Data Protection 
    Regulation (GDPR)~\cite{GDPR} and ePrivacy~\cite{ePrivacy}.
    
    \item \textbf{Journalists} - Regardless of 
    whether one takes the angle of the market structure of online 
    tracking, or conceptually facilitating the education of the consumers 
    on the issue, journalists will have enough data to derive insights from.

    \item \textbf{Web Developers} - Certain third-party scripts that web developers may 
    add to their sites, have the capacity of loading other third parties, which 
    the web developer may or may not know about. This, for instance, is the typical 
    behaviour of ad networks like DoubleClick.
    Web developers can use \wtm to keep an eye on the extent to
    which they retain control over third parties loaded, which will be important
    in the context of GDPR compliance~\cite{GDPR}. Furthermore, not doing so
    can often have undesired consequences.
    
    \item \textbf{Block-list maintainers} - Can benefit from the automatic
    discovery of trackers, and can easily use the open source data
    to generate block lists\footnote{\url{https://whotracks.me/blog/generating_adblocker_filters.html}}.
    
    \item \textbf{Everyone} - Can build understanding of their exposure to 
    tracking by learning about the tracking landscape on their favourite
    websites and read the educational resources in the \wtm blog.
    
\end{itemize}

\section{Summary \& Conclusions}

As the line between the physical and online lives becomes 
more blurred, we believe online privacy will gain the attention of academics, 
regulators, media and users at large. In the context of paving the 
way for a constructive approach to dealing with online tracking,
we open source the \wtm data, which we plan to maintain, 
and update on a monthly basis.

This paper, and the living representation of it: \wtm, 
contribute to the body of research, and public sphere more 
broadly, in the following ways:

\begin{itemize}[noitemsep,topsep=0pt]
    \item \textbf{Largest dataset on web tracking} to our knowledge. This 
    assists researchers, regulators, journalists, web developers and users
    in developing efficient tools, devising policies and running awareness campaigns
    to address the negative externalities tracking introduces.  

    \item \textbf{Longitudinal data}: While snapshots of data are necessary, 
    in the non-transparent environment of online tracking, for the 
    purpose of monitoring, it is also important to have have longitudinal 
    data. \wtm open sources data from the longest measurement of 
    web tracking to date.

    \item \textbf{Measuring without platform-side-effects}: The data is 
    generated by the behaviour of real users, which means the data 
    is not prone to effects introduced by the measuring platform.

    \item \textbf{Human-Machine cooperation}: A significant amount of browser
    privacy tools, rely on publicly maintained block lists. \wtm data
    contains trackers profiled algorithmically, as presented in \cite{DBLP:conf/www/YuMMP16}.
    Assisting the maintenance of blocklists, the community can focus
    on the accuracy of demographic data of the identified trackers, thus
    collectively improving transparency.

    \item \textbf{Measuring the effects of regulation}: The longitudinal nature of
    the data, enables users of \wtm to measure the effects of regulation on the 
    tracking landscape. An example of such application will be the measuring of
    effects the implementation of the General Data Protection Regulation (GDPR), 
    in May 2018 will have on tracking practices.

\end{itemize}

Given increasing concern over the data collected by often nameless third-parties 
across the web, and consumers' struggles to keep control of their data trails, 
more transparency, accountability and monitoring is required in the ecosystem. 
This work represents a step-change in the quantity and depth of information
available to those who wish to push for a healthier web.


%
\bibliographystyle{abbrv}
\bibliography{privacy}  
\end{document}